\documentstyle[12pt]{article}
\input epsfig

\begin{document}
\baselineskip=20.5pt
\def\beqra{\begin{eqnarray}} \def\eeqra{\end{eqnarray}}
\def\beqast{\begin{eqnarray*}} \def\eeqast{\end{eqnarray*}}
\def\beq{\begin{equation}}      \def\eeq{\end{equation}}
\def\be{\begin{enumerate}}   \def\ee{\end{enumerate}}

\def\fnote#1#2{\begingroup\def\thefootnote{#1}\footnote{#2}\addtocounter
{footnote}{-1}\endgroup}

\def\gam{\gamma}
\def\Gam{\Gamma}
\def\la{\lambda}
\def\eps{\epsilon}
\def\La{\Lambda}
\def\si{\sigma}
\def\Si{\Sigma}
\def\al{\alpha}
\def\Tha{\Theta}
\def\tha{\theta}
\def\vphi{\varphi}
\def\del{\delta}
\def\Del{\Delta}
\def\ab{\alpha\beta}
\def\om{\omega}
\def\Om{\Omega}
\def\mn{\mu\nu}
\def\mun{^{\mu}{}_{\nu}}
\def\kap{\kappa}
\def\rsi{\rho\sigma}
\def\beal{\beta\alpha}
\def\til{\tilde}
\def\rta{\rightarrow}
\def\eqv{\equiv}
\def\nab{\nabla}
\def\pa{\partial}
\def\sit{\tilde\sigma}
\def\ul{\underline}
\def\indt{\parindent2.5em}
\def\nd{\noindent}
\def\rsi{\rho\sigma}
\def\beal{\beta\alpha}
\def\caa{{\cal A}}
\def\cb{{\cal B}}
\def\cac{{\cal C}}
\def\cd{{\cal D}}
\def\ce{{\cal E}}
\def\cf{{\cal F}}
\def\cg{{\cal G}}
\def\cah{{\cal H}}
\def\ci{{\cal I}}
\def\cj{{\cal{J}}}
\def\ck{{\cal K}}
\def\cl{{\cal L}}
\def\cm{{\cal M}}
\def\cn{{\cal N}}
\def\cO{{\cal O}}
\def\cp{{\cal P}}
\def\car{{\cal R}}
\def\cs{{\cal S}}
\def\ct{{\cal{T}}}
\def\cu{{\cal{U}}}
\def\cv{{\cal{V}}}
\def\cw{{\cal{W}}}
\def\cx{{\cal{X}}}
\def\cy{{\cal{Y}}}
\def\cz{{\cal{Z}}}
\def\asymptotic{{_{\stackrel{\displaystyle\longrightarrow}
{x\rightarrow\pm\infty}}\,\, }} 
\def\asymptext{\raisebox{.6ex}{${_{\stackrel{\displaystyle\longrightarrow}
{x\rightarrow\pm\infty}}\,\, }$}} 
\def\asymptoticp{{_{\stackrel{\displaystyle\longrightarrow}
{x\rightarrow +\infty}}\,\, }} 
\def\asymptoticm{{_{\stackrel{\displaystyle\longrightarrow}
{x\rightarrow -\infty}}\,\, }} 

\def\raisenot{\raise .5mm\hbox{/}}
\def\nota{\ \hbox{{$a$}\kern-.49em\hbox{/}}}
\def\notA{\hbox{{$A$}\kern-.54em\hbox{\raisenot}}}
\def\notb{\ \hbox{{$b$}\kern-.47em\hbox{/}}}
\def\notB{\ \hbox{{$B$}\kern-.60em\hbox{\raisenot}}}
\def\notc{\ \hbox{{$c$}\kern-.45em\hbox{/}}}
\def\notd{\ \hbox{{$d$}\kern-.53em\hbox{/}}}
\def\notbd{\ \hbox{{$D$}\kern-.61em\hbox{\raisenot}}} 
\def\note{\ \hbox{{$e$}\kern-.47em\hbox{/}}}
\def\notk{\ \hbox{{$k$}\kern-.51em\hbox{/}}}
\def\notp{\ \hbox{{$p$}\kern-.43em\hbox{/}}}
\def\notq{\ \hbox{{$q$}\kern-.47em\hbox{/}}}
\def\notW{\ \hbox{{$W$}\kern-.75em\hbox{\raisenot}}}
\def\notz{\ \hbox{{$Z$}\kern-.61em\hbox{\raisenot}}}
\def\notpa{\hbox{{$\partial$}\kern-.54em\hbox{\raisenot}}}

\def\fo{\hbox{{1}\kern-.25em\hbox{l}}}  
\def\rf#1{$^{#1}$}
\def\bx{\Box}
\def\tr{{\rm Tr}}
\def\rmtr{{\rm tr}}
\def\dgg{\dagger}
\def\lag{\langle}
\def\rag{\rangle}
\def\bmid{\big|}
\def\vlap{\overrightarrow{\La p}} 
\def\lrta{\longrightarrow} \def\lrar{\raisebox{.8ex}{$\longrightarrow$}}
\def\ON{{\cal O}(N)}
\def\UN{{\cal U}(N)}
\def\bdPh{\mbox{\boldmath{$\dot{\!\Phi}$}}}
\def\bPh{\mbox{\boldmath{$\Phi$}}}
\def\bPhs{\bPh^2}
\def\sef{S_{eff}[\sigma,\pi]}
\def\sigx{\sigma(x)}
\def\pix{\pi(x)}
\def\bph{\mbox{\boldmath{$\phi$}}}
\def\bphs{\bph^2}
\def\ex{\BM{x}}
\def\exs{\ex^2}
\def\xdot{\dot{\!\ex}}
\def\y{\BM{y}}
\def\ys{\y^2}
\def\ydot{\dot{\!\y}}
\def\pat{\pa_t}
\def\pax{\pa_x}

\renewcommand{\thesection}{\arabic{section}}
\renewcommand{\theequation}{\thesection.\arabic{equation}}


\vspace*{.2in}
\begin{center}
  \Large{\sc Generalized Supersymmetric Quantum Mechanics and 
Reflectionless Fermion Bags in $1+1$ Dimensions}\\
\normalsize
\vspace{15pt}
\begin{center}
{\bf Joshua Feinberg$^{a~*}$ \& A. Zee$^{b}$
\fnote{*}{{\it e-mail addresses: joshua@physics.technion.ac.il, 
zee@itp.ucsb.edu}}}
\end{center}
\vskip 2mm
\begin{center}
$^{a)}${Physics Department,}\\
{University of Haifa at Oranim, Tivon 36006, Israel\fnote{**}{{\it permanent address}}}\\
{and}\\
{Physics Department,}\\
{Technion, Israel Institute of Technology, Haifa 32000, Israel}\\
\vskip 2mm
$^{b)}${Institute for Theoretical Physics}\\
{University of California\\ Santa Barbara, CA 93106, USA}\\
\end{center}
\vspace{.3cm}
\end{center}

\begin{minipage}{6.3in}
{\abstract~~~~~
We study static fermion bags in the $1+1$ dimensional Gross-Neveu and
Nambu-Jona-Lasinio models. It has been known, from the work of
Dashen, Hasslacher and Neveu (DHN), followed by Shei's work, in the
1970's, that the self-consistent static fermion bags in these models 
are reflectionless. The works of DHN and of Shei were based on
inverse scattering theory. Several years ago, we offered an alternative
argument to establish the reflectionless nature of these fermion bags,
which was based on analysis of the spatial asymptotic behavior of the
resolvent of the Dirac operator in the background of a static bag,
subjected to the appropriate boundary conditions. We also calculated 
the masses of fermion bags based on the resolvent and the Gelfand-Dikii 
identity. Based on arguments taken from a certain generalized one dimensional
supersymmetric quantum mechanics, which underlies the spectral theory of
these Dirac operators, we now realize that our analysis of the asymptotic
behavior of the resolvent was incomplete. We offer here a critique of our
asymptotic argument.}
\end{minipage}

\vspace{10pt}

\vfill
\pagebreak

\setcounter{page}{1}

\section{Introduction}

Many years ago, Dashen, Hasslacher and Neveu (DHN) \cite{dhn}, and following 
them Shei \cite{shei}, used inverse scattering analysis \cite{faddeev} to find 
static fermion-bag \cite{sphericalbag, shellbag} soliton solutions to the 
large-$N$ saddle point equations of the Gross-Neveu (GN) \cite{gn} and of the 
$1+1$ dimensional, multi-flavor Nambu-Jona-Lasinio (NJL) \cite{njl} models.
In the GN model, with its discrete chiral symmetry, a topological soliton,
the so called Callan-Coleman-Gross-Zee (CCGZ) kink \cite{ccgz}, was 
discovered prior to the work of DHN.

One version of writing the action of the $1+1$ dimensional NJL model is 
\beq
S=\int d^2x\,\left\{\sum_{a=1}^N\, \bar\psi_a\,\left[i\notpa-(\si+i\pi\gam_5)
\right]\,\psi_a 
-{1\over 2g^2}\,(\si^2+\pi^2)\right\}\,,
\label{auxiliary}
\eeq
where the $\psi_a\,(a=1,\ldots,N)$ are $N$ flavors of massless Dirac 
fermions, with Yukawa couplings to the scalar and pseudoscalar auxiliary 
fields $\si(x), \pi(x)$\footnote{The fermion 
bag solitons in these models arise, as is well known, at the level of the 
effective action, after integrating the fermions out, and not at the level of 
the action (\ref{auxiliary}).}.

The remarkable discovery DHN made was that all these static bag configurations
were {\em reflectionless}. More precisely, the static $\si (x)$ and $\pi (x)$ 
configurations, that solve the saddle point equations of the NJL model,
are such that the the Dirac equation 
\beq\label{diraceq}
\left[i\notpa-\si (x)-i\pi (x) \gam_5\right]\,\psi (x) = 0
\eeq
in these backgrounds has scattering solutions, whose reflection
amplitudes at momentum $k$ vanish identically for {\bf all} values of $k$. 
In other words, a fermion wave packet impinging on one side of the potential 
well $\si (x)+i\pi (x) \gam_5, $ will be totally transmitted through the well 
(up to phase shifts, of course).

We note in passing that besides their role in soliton theory 
\cite{faddeev, jg}, 
reflectionless potentials appear in other diverse areas of theoretical 
physics \cite{rosner, blackhole, zwiebach}. For a review, which discusses 
reflectionless potentials (among other things) in the context of 
supersymmetric quantum mechanics, see \cite{cooper}. 

Since the works of DHN and of Shei, these fermion bags were discussed in the 
literature several other times, using alternative methods \cite{others}. 
For a recent review on these and related matters, see \cite{thies}. 
Very recently, static chiral fermion bag solitons \cite{jaffe} in a $1+1$ 
dimensional model, as well as non-chiral (real scalar) fermion bag solitons 
\cite{bashinsky}, were discussed, in which the scalar field that couples to 
the fermions was dynamical already at the classical level 
(unlike the auxiliary fields $\si$ and $\pi$ in (\ref{auxiliary})).

In many of these treatments, one solves the variational, saddle point
equations by performing mode summations over energies and phase shifts. 
An alternative to such summations is to solve the saddle point equations 
by manipulating the resolvent of the Dirac operator as a whole, with the 
help of simple tools from Sturm-Liouville operator theory. The resolvent 
of the Dirac operator takes care of mode summation automatically.

Some time ago, one of us had developed such an alternative to the inverse 
scattering method, which was based on the Gel'fand-Dikii (GD) identity 
\cite{gd} (an identity obeyed by the diagonal resolvent of one-dimensional 
Schr\"odinger operators)\footnote{For a simple derivation of the GD identity,
see \cite{josh2, fz}.}, to study fermion bags in the GN model\cite{josh1} 
as well as other problems \cite{josh2}. 
That method was later applied by us to study fermion bags in the NJL model 
\cite{fz} and in the massive GN model \cite{massivegn}. Similar ideas were 
later used in \cite{stone} to calculate the free energy of inhomogeneous 
superconductors.

Application of this method in \cite{josh1} and in \cite{fz} reproduced the 
static bag results of DHN and of Shei in what seems to be a simpler manner 
than in the inverse scattering formalism. In \cite{fz}, we followed 
the method introduced in \cite{josh1, josh2}, and simply wrote down
an efficient, parameter dependent, ansatz for the diagonal resolvent of the 
Dirac operator in a static $\sigx$, $\pix$ background. Construction of that
ansatz was based on simple dimensional analysis, and on the Gelfand-Dikii
identity. Nowhere in the construction, did we use the theory of 
reflectionless potentials. With the help of that ansatz, we were able to 
reproduce in \cite{fz} Shei's inverse scattering results, in a similar 
manner to the reproduction of DHN's results in \cite{josh1}.

In addition to rederivation of bag profiles, masses and quantum numbers 
found by DHN and Shei, we tried 
in \cite{fz} to explain the reflectionless property of the static 
background in simple terms, by studying the expectation value of the 
fermion current $\langle j^1 (x) \rangle = 
\langle \bar\psi (x) \gam^1 \psi (x) \rangle $ in a given background 
$\sigma (x)  + i\gam_5 \pi (x)$, at spatial asymptotics.  However, after 
careful reexamination, we now realize that the part of the 
analysis in \cite{fz} on the reflectionless
nature of the background was incomplete.\footnote{We thank R. Jaffe and 
N. Graham for useful correspondence on this point.} We realized this with 
the help of a certain version of generalized one dimensional 
supersymmetric (SUSY) quantum mechanics \cite{generalizedsusy}, that 
underlies the spectral theory of the Dirac operator in (\ref{diraceq}).

This paper offers critique of our asymptotic argument from \cite{fz}.
The rest of the results in \cite{fz}, namely, bag profiles etc., remain 
intact, and will not be discussed here.

Before discussing this issue in detail, and in order to set our notations,
let us recall some basic facts about dynamics of the NJL model:

The partition function associated with (\ref{auxiliary}) 
is\footnote{From this point to the end of this 
paper flavor indices are usually suppressed. Thus $i\bar\psi\notpa\psi$ 
should be 
understood as $\displaystyle{i\sum^N_{a=1} \bar\psi_a \notpa\psi_a}$.  
Similarly $\bar\psi\Gamma\psi$ stands for $\displaystyle{\sum^N_{a=1} 
\,\bar\psi_a\Gamma\psi_a}$, where $\Gamma=1,\gam_5$.}
\beq
\cz=\int\,\cd\si\,\cd\pi\,\cd\bar\psi\,\cd\psi \,\exp i\, 
\int\,d^2x\left\{\bar\psi
\left[i\notpa-\left(\si+i\pi\gam_5\right)\right]\psi-{1\over 
2g^2}\,\left(\si^2+\pi^2\right)\right\}
\label{partition}
\eeq
Integrating over the grassmannian variables leads to 
$\cz=\int\,\cd\si\,\cd\pi\,\exp \{iS_{eff}[\si,\pi]\}$
where the bare effective action is
\beq
S_{eff}[\si,\pi] =-{1\over 2g^2}\int\, d^2x 
\,\left(\si^2+\pi^2\right)-iN\, 
\tr\log\left[i\notpa-\left(\si+i\pi\gam_5\right)\right]
\label{effective}
\eeq
and the trace is taken over both functional and Dirac indices. 

This theory has been studied in the limit 
$N\rightarrow\infty$ with $Ng^2$ held fixed\cite{gn}. In this limit 
(\ref{partition}) is governed by saddle points of (\ref{effective}) 
and the small fluctuations around them. The most general saddle point
condition reads

\beqra
{\del S_{\em eff}\over \del \si\left(x,t\right)}  &=&
-{\si\left(x,t\right)\over g^2} + iN ~{\rm tr} \left[~~~~~ \langle x,t | 
{1\over i\notpa
-\left(\si + i\pi\gam_5\right)} | x,t \rangle \right]= 0
\nonumber\\{}\nonumber\\
{\del S_{\em eff}\over \del \pi\left(x,t\right)}  &=&
-{\pi\left(x,t\right)\over g^2} - ~N~ {\rm tr} \left[~\gam_5~\langle x,t 
| {1\over i\notpa
-\left(\si + i\pi\gam_5\right)} | x,t \rangle~\right] = 0\,.
\label{saddle}
\eeqra

In particular, the non-perturbative vacuum of (\ref{auxiliary}) is 
governed 
by the simplest large $N$ saddle points of the path integral associated 
with 
it, where the composite scalar operator $\bar\psi\psi$ and the 
pseudoscalar operator $i\bar\psi\gam_5\psi$ develop space-time 
independent expectation values.  

These saddle points are extrema of the effective potential $V_{eff}$ 
associated with (\ref{auxiliary}), namely, the value of  $-S_{eff}$ for 
space-time independent $\si,\pi$ configurations per unit time per unit 
length.  The effective potential $V_{eff}$ 
depends only on the combination $\rho^2=\si^2+\pi^2$ as a result of 
chiral symmetry. $V_{eff}$ has a minimum as a function of $\rho$ at  
$\rho = m \neq 0$ 
that is fixed by the (bare) gap equation\cite{gn}
\beq
-m + iNg^2\,{\rm tr}\int
{d^2k\over\left(2\pi\right)^2}{1\over\notk-m}
= 0
\label{bgap}
\eeq
which yields the dynamical mass
\beq
m = \Lambda\,e^{-{\pi\over Ng^2\left(\Lambda\right)}}\,.
\label{mass}
\eeq
Here $ \Lambda$ is an ultraviolet cutoff. The mass $m$ must be a 
renormalization group invariant. Thus, the model is asymptotically 
free. We can get rid of the cutoff at the price of introducing an 
arbitrary 
renormalization scale $\mu$. The renormalized
coupling $g_R\left(\mu\right)$ and the cut-off dependent bare
coupling are then related through $ \Lambda\,e^{-{\pi\over 
Ng^2\left(\Lambda\right)}} = 
\mu\,e^{1-{\pi\over Ng_R^2\left(\mu\right)}}$ in a convention where 
$Ng_R^2\left(m\right) = {1\over\pi}$. Trading the dimensionless coupling 
$g_R^2$ for the dynamical mass scale $m$ represents the well known 
phenomenon of dimensional transmutation. 

The vacuum manifold of (\ref{auxiliary}) is therefore a circle 
$\rho=m$ in the $\si,\pi$ plane, and the equivalent vacua are 
parametrized by the chiral angle $\theta={\rm arctan} {\pi\over\sigma}$. 
Therefore, small fluctuations of the Dirac fields around the vacuum 
manifold develop dynamical 
chiral mass $m\,{\rm exp} (i\theta \gam_5)$.

Note in passing that the massless fluctuations of $\theta$ along the 
vacuum manifold decouple from the spectrum \cite{decouple}
so that the axial $U(1)$ symmetry does not break dynamically in this two 
dimensional model \cite{coleman}, in accordance with the 
Coleman-Mermin-Wagner theorem.

Non-trivial excitations of the vacuum, on the other hand, are described 
semiclassically by large $N$ saddle points of the path integral over 
(\ref{auxiliary}) at which $\si$ and $\pi$ develop space-time dependent 
expectation values\cite{cjt,jg}. These expectation values are the 
space-time dependent solution of (\ref{saddle}). Saddle points of this 
type are important 
also in discussing the large order behavior\cite{bh,dev} of the 
${1\over N}$ expansion of the path integral over (\ref{auxiliary}).

These saddle points describe sectors of (\ref{auxiliary}) that include 
scattering states of the (dynamically massive) fermions in 
(\ref{auxiliary}), as well as a rich collection of bound states thereof. 

These bound states result from the strong infrared interactions, 
which polarize the vacuum inhomogeneously, causing 
the composite scalar $\bar\psi\psi$ and pseudoscalar 
$i\bar\psi\gam_5\psi$ 
fileds to form finite action space-time dependent condensates. These 
condensates are stable because of the binding energy released by the 
trapped fermions and therefore cannot form without such binding. This
description agrees with the general physical picture drawn in 
\cite{mackenzie}. 
We may regard these condensates as one dimensional chiral 
bags \cite{sphericalbag,shellbag} that trap the original fermions 
(``quarks") into 
stable finite action extended entities (``hadrons").

If we set $\pix$ in (\ref{auxiliary}) to be identically zero, we recover 
the Gross-Neveu model, defined by
\beq
S_{GN}=\int d^2x\,\left\{\bar\psi\left[i\notpa-\si\right]\psi-{\si^2 
\over 
2g^2}\right\}\,.
\label{GN}
\eeq

In spite of their similarities, these two field theories are quite
different, as is well-known from the field theoretic literature of the
seventies. The crucial difference is that the Gross-Neveu model 
possesses a discrete symmetry, $\sigma\rightarrow -\sigma$, rather than 
the continuous axial $U(1)$ symmetry 
$\si + i\gam_5 \pi \rightarrow e^{-i\gam_5\alpha}(\si + i\gam_5 \pi)$ in the 
NJL model (\ref{auxiliary}).
This discrete symmetry
is dynamically broken by the non-perturbative vacuum, and thus there is a 
kink solution \cite{ccgz,dhn,josh1}, the CCGZ kink mentioned above, $\sigx = 
m\,{\rm tanh}(mx)$, interpolating between $\pm m$ at  $x= \pm \infty$ 
respectively. Therefore, topology insures the stability of these kinks.

In contrast, the NJL model, with its continuous symmetry, does 
not have a topologically stable soliton solution. The solitons arising 
in the NJL model can only be stabilized by binding fermions, namely, 
stability of fermion bags in the NJL model is not due to topology, but to 
dynamics.

The rest of the paper is organized as follows: 
In Section 2 we review the results of \cite{generalizedsusy}. We 
study the resolvent of the Dirac operator in a given static
$\sigx + i\gam_5\pix$ background. The Dirac equation in any such 
background has special properties. In fact, we show that it is equivalent 
to a pair of two isospectral Sturm-Liouville equations in one dimension, 
which generalize the well known one-dimensional supersymmetric 
quantum mechanics. We use this generalized supersymmetry to express all 
four entries of the space-diagonal Dirac resolvent 
(i.e., the resolvent evaluated at coincident spatial coordinates) in terms 
of a single function. As a result, we can prove that each frequency mode of 
the spatial current $\langle \bar\psi (x) \gam^1 \psi (x)\rangle$ 
vanishes identically, contrary to the argument we made in 
\cite{fz}. The findings of Section 2 are then used in Section 3 to 
simplify the saddle point equations (\ref{saddle}). 
We then study the spatial asymptotic behavior of the simplified equations. 
We use the spatial asymptotic expression of the resolvent of the Dirac 
operator (summarized in the Appendinx) to generate an asymptotic 
expansion of the quantities ${\del S_{\em eff}\over\del \si\left(x,t\right)}$
and $\del S_{\em eff}\over \del \pi\left(x,t\right)$, evaluated on 
a static background $(\sigx,\pix)$ (consistent with the physical boundary 
conditions at spatial infinity). We prove that these asymptotic expansions
vanish term by term to any power in $1/x$, for {\em any} 
static $\sigx$ and $\pix$ that are consistent with the physical boundary 
conditions, and not just for reflectionless backgrounds, as we have claimed 
in \cite{fz}. In the Appendix we 
recall the asymptotic behavior of resolvents of Sturm-Liouville operators 
and use them to derive the asymptotic behavior of the resolvent of the Dirac 
operator in a static bag background. 
\pagebreak

\section{Resolvent of the Dirac Operator With Static Background Fields }
\setcounter{equation}{0}

As was explained in the introduction, we are interested in static space 
dependent solutions of the extremum condition on $S_{\it eff}$. To this 
end 
we need to invert the Dirac operator 
\beq
D\equiv i\notpa-(\sigx+i\pix\gam_5)
\label{dirac}
\eeq
in a given background of static field configurations $\sigx$ and $\pix$. 
In particular, we have to find the diagonal resolvent of (\ref{dirac}) in 
that background. We stress that inverting (\ref{dirac}) has nothing to do 
with the large $N$ approximation, and consequently our results in this 
section are valid for any value of $N$. For example, our results may be of 
use in generalizations of supersymmetric quantum mechanics. 

For the usual physical reasons, we set boundary conditions on our static 
background fields such that $\sigx$ and $\pix$ start from a point on the 
vacuum manifold $\si^2 + \pi^2 = m^2$ at $x=-\infty$, wander around in the
$\si-\pi$ plane, and then relax back to another point on the vacuum manifold
at $x=+\infty$. Thus, we must have the asymptotic behavior 
\beqra
&&\si\asymptotic m{\rm cos}\theta_{\pm}\quad\quad ,\quad\quad 
\si'\asymptotic 0 
\nonumber\\{}\nonumber\\
&&\pi\asymptotic m{\rm sin}\theta_{\pm}\quad\quad , 
\quad\quad \pi'\asymptotic 0
\label{boundaryconditions}
\eeqra
where $\theta_{\pm}$ are the asymptotic chiral alignment angles. Only the 
difference $\theta_+ - \theta_-$ is meaningful, of course, and henceforth 
we use the axial $U(1)$ symmetry to set $\theta_- = 0$, such that $\si 
(-\infty)=m$ and $\pi (-\infty)=0$. We also omit the subscript from 
$\theta_+$
and denote it simply by $\theta$ from now on. As typical of solitonic
configurations, we expect, that $\sigx$ and $\pix$ tend to their asymptotic 
boundary values (\ref{boundaryconditions}) on the vacuum manifold at an 
exponential rate which is determined, essentially, by the mass gap $m$ of 
the model. It is in the background of such fields that we wish to invert 
(\ref{dirac}).

In this paper we use the Majorana
representation  
\beq\label{majorana}
\gam^0=\si_2\;,\; \gam^1=i\si_3\quad {\rm and} \quad 
\gam^5=-\gam^0\gam^1=\si_1
\eeq
for $\gam$ matrices. In this representation  (\ref{dirac}) becomes
\beq
D =\left(\begin{array}{cc} -\pa_x - \si & -i\omega - i\pi \\{}&{}\\ 
i\omega- i\pi & \pa_x - \si\end{array}\right)
=\left(\begin{array}{cc} -Q & -i\omega - i\pi \\{}&{}\\ 
i\omega- i\pi & -Q^\dgg\end{array}\right)\,,
\label{dirac1}
\eeq
where we introduced the pair of adjoint operators 
\beq\label{qqdagger}
Q = \sigx + \pa_x\,,\quad\quad Q^\dgg = \sigx - \pa_x\,.
\eeq
(To obtain (\ref{dirac1}), we have naturally transformed 
$i\notpa-(\sigx+i\pix\gam_5)$ to the $\om$ plane, since the 
background fields $\sigx, \pix$ are static.) 

Inverting (\ref{dirac1}) is achieved by solving  
\beq
\left(\begin{array}{cc} -Q & -i\omega - i\pix \\{}&{}\\ 
i\omega- i\pix & -Q^\dgg\end{array}\right)\cdot 
\left(\begin{array}{cc} a(x,y) &  b(x,y) \\{}&{}\\ c(x,y) & 
d(x,y)\end{array}\right)\,=\,-i{\bf 1}\del(x-y)
\label{greens}
\eeq
for the Green's function of (\ref{dirac1}) in a given background 
$\sigx,\pix$.
By dimensional analysis, we see that the quantities $a,b,c$ and 
$d$ are dimensionless.

\subsection{Generalized ``Supersymmetry'' in a Chiral Bag Background}
Interestingly, the spectral theory of the Dirac operator (\ref{dirac1})
is underlined by a certain generalized one dimensional supersymmetric 
quantum mechanics \cite{generalizedsusy}. This generalized supersymmetry
is very helpful in simplifying various calculations involving the Dirac 
operator and its resolvent. In the remaining part of this section, 
we review the discussion in \cite{generalizedsusy}.

The diagonal elements $a(x,y), ~d(x,y)$ in (\ref{greens}) may be 
expressed
in term of the off-diagonal elements as
\beq
a(x,y)={-i\over \omega-\pix}Q^\dgg c(x,y)\,,\quad\quad
d(x,y)={i\over \omega+\pix} Q b(x,y)
\label{ad}
\eeq
which in turn satisfy the second order partial differential equations
\beqra
&&\left[Q^\dgg {1\over \om+\pix}Q -(\om-\pix )\right]b(x,y)\,=
\nonumber\\&&{}\nonumber\\
&&-\pa_x\left[{\pa_x b(x,y)\over 
\omega+\pix}\right]+\left[\sigx^2+\pix^2-\si'(x)-\omega^2+{\sigx\pi'(x)
\over 
\omega+\pix}\right]{b(x,y)\over \omega+\pix}\,=\,~~\del(x-y)
\nonumber\\&&{}\nonumber\\
&&\left[Q {1\over \om-\pix}Q^\dgg -(\om+\pix )\right]c(x,y)\,=
\nonumber\\&&{}\nonumber\\
&&-\pa_x\left[{\pa_x c(x,y)\over 
\omega-\pix}\right]+\left[\sigx^2+\pix^2+\si'(x)-\omega^2+{\sigx\pi'(x)
\over 
\omega-\pix}\right]{c(x,y)\over 
\omega-\pix}\,=\,-\del(x-y)\,.\nonumber\\&&{}
\label{bc}
\eeqra

Thus, $b(x,y)$ and $-c(x,y)$ are simply the Green's functions of the 
corresponding second order Sturm-Liouville operators\footnote{Note that 
$\om$ plays here a dual role: in addition to its role as the spectral 
parameter (the $\om^2$ terms in (\ref{bcops})), it also appears as a
parameter in the definition of these operators-hence the explicit $\om$ 
dependence in our notations for these operators in (\ref{bcops}). However, 
in order to avoid notational cluttering, from now on we will denote these 
operators simply as $L_b$ and $L_c$.}
\beqra\label{bcops}
&& L_b(\om) b(x) = -\pa_x\left[{\pa_x b(x)\over 
\omega+\pix}\right]+\left[\sigx^2+\pix^2-\si'(x)-\omega^2+{\sigx\pi'(x)
\over 
\omega+\pix}\right]{b(x)\over \omega+\pix}
\nonumber\\&&{}\nonumber\\
&& L_c(\om) c(x) = -\pa_x\left[{\pa_x c(x)\over 
\omega-\pix}\right]+\left[\sigx^2+\pix^2+\si'(x)-\omega^2+{\sigx\pi'(x)
\over 
\omega-\pix}\right]{c(x)\over 
\omega-\pix}\nonumber\\&&{}
\eeqra
in (\ref{bc}), namely, 
\beqra
b(x,y)&=&{\theta\left(x-y\right)b_2(x)b_1(y)+\theta\left(y-x\right)b_2(y)
b_1(x) \over W_b}
\nonumber\\{}\nonumber\\
c(x,y)&=&-{\theta\left(x-y\right)c_2(x)c_1(y)+\theta\left(y-x\right)c_2(y
)c_1(x)\over W_c}\,.
\label{bcexpression}
\eeqra
Here $\{b_1(x), b_2(x)\}$ and $\{c_1(x), c_2(x)\}$ are pairs of independent 
fundamental solutions of the two equations $L_b b(x)=0$ and $L_c c(x)=0$, 
subjected to the 
boundary conditions 
\beq\label{planewaves}
b_1(x)\,,c_1(x) \asymptoticm A_{b,c}^{(1)}(k)e^{-ikx} \quad\quad ,\quad\quad 
b_2(x)\,,c_2(x)\asymptoticp A_{b,c}(k)^{(2)}e^{ikx}
\eeq
with some possibly $k$ dependent coefficients $A_{b,c}^{(1)}(k), 
A_{b,c}^{(2)}(k)$ and with\footnote{We see that if ${\rm Im}k>0$, $b_1$ 
and $c_1$ decay exponentially to the left, and $b_2$ and $c_2$ decay to 
the right. Thus, if ${\rm Im}k>0$, both  $b(x,y)$ and $c(x,y)$ decay as 
$|x-y|$ tends to infinity.} 
\beq\label{kmom}
k=\sqrt{\om^2 -m^2}\,,\quad {\rm Im}k\geq 0\,.
\eeq
The purpose of introducing the (yet unspecified) 
coefficients $A_{b,c}^{(1)}(k), A_{b,c}^{(2)}(k)$ 
will become clear following Eqs. (\ref{btoc}) and (\ref{ctob}). 
The boundary conditions (\ref{planewaves}) are consistent, of course, with 
the asymptotic behavior (\ref{boundaryconditions}) of $\si$ and $\pi$ due to 
which both $L_b$ and $L_c$ tend to a free particle hamiltonian 
$[-\pa_x^2 + m^2-\om^2]$ as $x\rightarrow\pm\infty$.

The wronskians of these pairs of solutions are 
\beqra\label{wronskian}
&& W_b(k) ={b_2(x)b_1^{'}(x)-b_1(x)b_2^{'}(x)\over \omega+\pix}
\nonumber\\&&{}\nonumber\\
&& W_c(k) = {c_2(x)c_1^{'}(x)-c_1(x)c_2^{'}(x)\over \omega-\pix}
\nonumber\\&&{}
\eeqra
As is well known, $W_b(k)$ and $W_c (k)$ are independent of $x$.

Note in passing that the canonical asymptotic behavior assumed in 
the scattering theory of the operators $L_b$ and $L_c$ corresponds to setting 
$A_{b,c}^{(1)} = A_{b,c}^{(2)} = 1$ in (\ref{planewaves}). Thus, 
the wronskians in (\ref{wronskian}) are {\em not} the canonical wronskians 
used in scattering theory. As is well known in the literature\cite{faddeev},
the {\em canonical} wronskians are proportional 
(with a $k$ independent coefficient) to $k/t(k)$, where $t(k)$ is 
the transmission amplitude of the corresponding 
operator $L_b$ or $L_c$. Thus, on top of the well-known features of $t(k)$, 
the wronskians in (\ref{wronskian}) will have additional spurious 
$k$-dependence coming from the amplitudes 
$A_{b,c}^{(1)}(k), A_{b,c}^{(2)}(k)$ in (\ref{planewaves}).

Substituting the expressions (\ref{bcexpression}) for the off-diagonal 
entries $b(x,y)$ and $c(x,y)$ into (\ref{ad}), we obtain the appropriate 
expressions for the diagonal entries $a(x,y)$ and $d(x,y)$. We do not bother
to write these expressions here. It is useful however to note, that despite 
the $\pax$'s in the $Q$ operators in (\ref{ad}), that act on the 
step functions in (\ref{bcexpression}), neither $a(x,y)$ nor $d(x,y)$ contain 
pieces proportional to $\del(x-y)$\,. Such pieces cancel one another due 
to the symmetry of (\ref{bcexpression}) under $x\leftrightarrow y$.

We will now prove that the spectra of the operators $L_b$ and $L_c$ 
are essentially the same. Our proof is based on the fact that we can 
factorize the eigenvalue equations $L_b b(x)=0$ and $L_c c(x)=0$ as
\beqra\label{bcfactor}
&& {1\over \om-\pix}\,Q^\dgg\, {1\over \om +\pix}\, Q\, b = b
\nonumber\\&&{}\nonumber\\
&& {1\over \om+\pix}\,Q\, {1\over \om -\pix}\, Q^\dgg\, c = c\,,
\nonumber\\&&{}
\eeqra
as should be clear from (\ref{bc}) and (\ref{bcops}).

The factorized equations (\ref{bcfactor}) suggest the following map between
their solutions. Indeed, given that $L_b b(x) = 0$, then 
clearly
\beq\label{btoc}
c(x) = {1\over \om +\pix} Q\, b(x)
\eeq
is a solution of $L_c c(x) = 0$. Similarly, if $L_c c(x) =0$, then 
\beq\label{ctob}
b(x) = {1\over \om -\pix} Q^\dgg \,c(x)
\eeq
solves $L_b b(x) =0$.

Thus, in particular, given a pair $\{b_1(x), b_2(x)\}$ of independent 
fundamental solutions of $L_b b(x) =0$, we can obtain from it 
a pair $\{c_1(x), c_2(x)\}$ of independent 
fundamental solutions of $L_c c(x) =0$ by using (\ref{btoc}), and vice
versa. Therefore, with no loss of generality, we henceforth assume,
that the two pairs of independent fundamental solutions $\{b_1(x), b_2(x)\}$
and $\{c_1(x), c_2(x)\}$, are related by (\ref{btoc}) and (\ref{ctob}).

The coefficients 
$A_{b,c}^{(1)}(k), A_{b,c}^{(2)}(k)$ in (\ref{planewaves}) are to be adjusted
according to (\ref{btoc}) and (\ref{ctob}), and this was the purpose of 
introducing them in the first place. 

Thus, with no loss of generality, we may make the standard choice 
\beq\label{standardchoice}
A_{b}^{(1)} = A_{b}^{(2)} =1
\eeq 
in (\ref{planewaves}). The coefficients $A_{c}^{(1)}, A_{c}^{(2)}$ are 
then determined by (\ref{btoc}):
\beqra\label{Accoeffs}
A_c^{(1)} &=& {\si(-\infty)-ik\over \pi (-\infty) + \om}\nonumber\\
{}\nonumber\\
A_c^{(2)} &=& {\si(\infty)+ik\over \pi (\infty) + \om}\,. 
\eeqra

We note that these $b(x)\leftrightarrow c(x)$ mappings can break only if 
\beq\label{mapbreak}
Q\,b = 0 \quad\quad {\rm or}\quad\quad Q^\dgg\,c =0\,,
\eeq
for $b(x)$ or $c(x)$ that {\em solve} (\ref{bcfactor}).  Do such solutions 
exist? Let us assume, for example, that $Q\,b = 0$ and that $L_b b=0$. From 
the first equation in (\ref{bcfactor}) (or in (\ref{bc})), we see that 
this is possible 
if and only if $\om\pm\pix\equiv 0$, which clearly cannot hold if 
$\pax\pix\neq 0$. A similar argument holds for $Q^\dgg\,c =0$. 
Thus, if $\pax\pix\neq 0$, the mappings (\ref{btoc}) and 
(\ref{ctob}) are one-to-one. In particular, a bound state in $L_b$ 
implies a bound state in $L_c$ (at the same energy) and vice-versa.

An interesting related result concerns the wronskians $W_b$ and $W_c$. 
From (\ref{wronskian}), and from (\ref{btoc}) and (\ref{ctob}) it follows 
immediately that for pairs of independent fundamental solutions 
$\{b_1(x), b_2(x)\}$ and $\{c_1(x), c_2(x)\}$ we have 
\beq\label{wbwc}
W_c = {c_2\pax c_1 - c_1\pax c_2\over \om-\pix} = c_1b_2 -c_2b_1 = 
{b_2\pax b_1 - b_1\pax b_2\over \om+\pix} = W_b\,.
\eeq
The wronskians of pairs of independent fundamental solutions of $L_b$ and 
$L_c$, which are related via (\ref{btoc}) and (\ref{ctob}) are equal!

To summarize, if $\pax\pix\neq 0$, $L_b$ and $L_c$ have the same set of 
energy eigenvalues and their eigenfunctions are in one-to-one correspondence. 

If, however, $\pi=$const., then we are back to the familiar ``supersymmetric'' 
factorization
\beq\label{susyfactor}
Q^\dgg\,Q\, b = (\om^2-\pi^2)\,b\,, \quad\quad 
Q\, Q^\dgg\, c = (\om^2-\pi^2)\,c\,,
\eeq
and mappings 
\beq\label{susymap}
c(x) = {1\over \om +\pi} Q\, b(x)\,,\quad\quad 
b(x) = {1\over \om -\pi} Q^\dgg \,c(x)\,.
\eeq
As is well known from the literature on supersymmetric quantum mechanics, 
the mappings (\ref{susymap}) break down if either $Qb=0$ or 
$Q^\dgg c=0$, in which case the two operators $Q^\dgg Q$ and $QQ^\dgg$ are 
isospectral, but only up to a ``zero-mode'' (or rather, an 
$\om^2=\pi^2$ mode), which belongs to the spectrum of only one of the 
operators\footnote{This is true for short range 
decaying potentials on the whole real line. For periodic potentials both 
operators may have that $\om^2=\pi^2$ mode in their spectrum \cite{fd}. 
Strictly speaking, (to the best of our knowledge) only the case 
$\pi=0$ appears in the literature on supersymmetric quantum mechanics.}. 
The case $\pix\equiv 0$ brings us back to the GN model. Supersymmetric 
quantum mechanical considerations were quite useful in the study of
fermion bags in \cite{josh1}. 

The ``Witten index'' associated with the pair of isospectral operators 
$L_b$ and $L_c$, is always null for backgrounds 
in which $\pax\pix\neq 0$, since they are absolutely isospectral, 
and not only up to zero modes. There is no interesting topology associated
with spectral mismatches of $L_b$ and $L_c$. This is not surprising at all, 
since, as we have already stressed in the introduction, the NJL model, with 
its continuous axial symmetry, does not support topological solitons. This is
in contrast to the GN model, for which $\pi\equiv 0$, which contains 
topological kinks, whose topological charge is essentially the Witten index
of the pair of operators (\ref{susyfactor}).

We note in passing that isospectrality of $L_b$ and $L_c$ which we have 
just proved, is consistent with the $\gam_5$ symmetry of the system 
of equations in (\ref{greens}), which relates the resolvent of $D$ with that  
of $\tilde D = -\gam_5 D \gam_5 $. Due to this symmetry, we can map 
the pair of equations $L_b b(x,y) = \delta (x-y)$ and 
$L_c c(x,y) = -\delta (x-y)$ (Eqs. (\ref{bc})) on each other by 
\beq\label{bcflip}
b(x,y)\leftrightarrow -c(x,y)\quad {\rm together~ with}\quad  
(\si,\pi)\rightarrow (-\si,-\pi)\,.
\eeq 
(Note that under these 
reflections we also have $a(x,y)\leftrightarrow -d(x,y)$, as we can see 
from (\ref{ad}).) The reflection $(\si,\pi)\rightarrow (-\si,-\pi)$ just 
shifts both asymptotic chiral angles $\theta_\pm$ by the same amount 
$\pi$, and clearly does not change the physics. Since this reflection 
interchanges $b(x,y)$ and $c(x,y)$ without affecting the physics, these two 
objects must have the same singularities as functions of $\om$, consistent 
with isospectrality of $L_b$ and $L_c$.

\subsection{The Diagonal Resolvent}

Following \cite{josh2,fz} we define the diagonal resolvent 
$\langle x\,|iD^{-1} | x\,\rangle$ symmetrically as
\beqra
\langle x\,|-iD^{-1} | x\,\rangle &\equiv& \left(\begin{array}{cc} A(x) & 
B(x) \\{}&{}\\ C(x) & 
D(x)\end{array}\right)\nonumber\\{}\nonumber\\{}\nonumber\\
 &=& {1\over 2} \lim_{\epsilon\rightarrow 0+}\left(\begin{array}{cc} 
a(x,y) + a(y,x) &  b(x,y) + b(y,x)\\{}&{}\\ c(x,y) +c(y,x) & d(x,y) + 
d(y,x)\end{array}\right)_{y=x+\epsilon}
\label{diagonal}
\eeqra
Here $A(x)$ through $D(x)$ stand for the entries of the diagonal 
resolvent, which following (\ref{ad}) and (\ref{bcexpression}) have the 
compact representation\footnote{$A, B, C$ and $D$ are obviously functions 
of $\omega$ as well. For notational simplicity we suppress their explicit 
$\omega$ dependence.}
\beqra
B(x)&=&~~{b_1(x)b_2(x)\over W_b}\quad\quad , \quad\quad D(x)={i\over 
2}{\left[\pax+2\sigx\right]B\left(x\right)\over 
\omega+\pix}\,,\nonumber\\
C(x)&=&-{c_1(x)c_2(x)\over W_c}\quad\quad , \quad\quad A(x)={i\over 
2}{\left[\pax-2\sigx\right]C\left(x\right)\over \omega-\pix}\,.
\label{abcd}
\eeqra

We now use the generalized ``supersymmetry'' of the Dirac operator,
which we discussed in the previous subsection, to deduce some important 
properties of the functions $A(x)$ through $D(x)$.

From (\ref{abcd}) and from (\ref{qqdagger}) we we have 
$$ A(x) = {i\over 2}{\pax-2\sigx\over \omega-\pix}\left(-{c_1c_2\over W_c}
\right)={i\over 2W_c}{c_2 Q^\dgg c_1 + c_1 Q^\dgg c_2 \over \omega-\pix}\,.$$
Using (\ref{ctob}) first, and then (\ref{btoc}), we rewrite this expression as 
$$A(x) = {i\over 2W_c}(c_2b_1 + c_1b_2) = {i\over 2W_c}  
{b_1 Q b_2  + b_2 Q b_1 \over \omega+\pix}\,.$$
Then, using the fact that $W_c=W_b$ (Eq. (\ref{wbwc})) and (\ref{abcd}),
we rewrite the last expression as 
$$A(x) = {i\over 2}{\pax + 2\si \over \omega+\pix}\left({b_1b_2\over W_b}
\right) = {i\over 2}{(\pax + 2\si)B(x) \over \omega+\pix}\,.$$
Thus, finally, 
\beq\label{ADeq}
A(x)=D(x)\,.
\eeq
Supersymmetry renders the diagonal elements $A$ and $D$ equal.

Due to (\ref{abcd}), $A=D$ is also a first order differential equation
relating $B$ and $C$. We can also relate the off diagonal elements 
$B$ and $C$ to each other more directly.
From (\ref{abcd}) and from (\ref{btoc}) we find
\beqra\label{cfromb}
C(x) = -{c_1c_2\over W_c} = -{(Qb_1)(Qb_2)\over (\om+\pi )^2 W_c}\,.
\eeqra
After some algebra, and using (\ref{wbwc}), we can rewrite this as 
$$ -(\om+\pi )^2 C = \si^2 B + \si B' + {b_1' b_2' \over W_b}$$
The combination $b_1'b_2'/W_b$ appears in $B''=(b_1b_2/W_b)''$. 
After using $L_b b_{1,2}=0$ to eliminate $b_1''$ and $b_2''$ from $B''$,
we find
$${b_1'b_2'\over W_b} = {1\over 2}B'' - {\pi'B'\over 2(\om+\pi)} - 
\left(\si^2 +\pi^2 - \si' -\om^2 + {\si\pi'\over \om + \pi}\right)B$$
Thus, finally, we have 
\beq\label{BtoCrel}
-(\om+\pi )^2 C = {1\over 2}B'' + \left( \si - {\pi'
\over 2(\om+\pi)}\right)B' - \left(\pi^2 - \si' -\om^2 + 
{\si\pi'\over \om + \pi}\right) B\,.
\eeq
In a similar manner we can prove that 
\beq\label{CtoBrel}
(\om-\pi )^2 B = -{1\over 2}C'' + \left( \si - {\pi'
\over 2(\om-\pi)}\right)C' + \left(\pi^2 + \si' -\om^2 + 
{\si\pi'\over \om - \pi}\right) C\,.
\eeq
We can simplify (\ref{BtoCrel}) and (\ref{CtoBrel}) further. 
After some algebra, and using (\ref{abcd}) we arrive at 
\beqra\label{BCfinal}
C(x) &=& {i\over \om+\pix}\,\pax D(x) - {\om - \pix \over \om + \pix}
\,B(x)\nonumber\\{}\nonumber\\
B(x) &=& {i\over \om - \pix}\,\pax A(x) - {\om + \pix \over \om - \pix}
\,C(x)\,.
\eeqra
Supersymmetry, namely, isospectrality of $L_b$ and $L_c$, enables us 
to relate the diagonal resolvents of these operators, $B$ and $C$, to each 
other. 

Thus, we can use (\ref{abcd}), (\ref{ADeq}) and (\ref{BCfinal})
to eliminate three of the entries of the diagonal resolvent in (\ref{abcd}),
in terms of the fourth. 

Note that the two relations in (\ref{BCfinal}) transform 
into each other under 
\beq\label{diagBCflip}
B\leftrightarrow -C\quad {\rm simultaneously~ with}\quad 
(\si,\pi)\rightarrow (-\si,-\pi)\,,
\eeq
in consistency with (\ref{bcflip}). 
The relations in (\ref{BCfinal}) are linear and homogeneous, with 
coefficients that for $\pax\pix\neq 0$ do not introduce additional 
singularities in the $\om$ plane. Thus, we see, once more, that 
$B$ and $C$ have the same singularities in the $\omega$ plane. We refer the 
reader to Section 4 in \cite{fz} for concrete examples of such resolvents.

The case $\pix\equiv 0$ brings us back to the GN model. In the GN model, our 
$B$ and $C$, coincide, respectively, with $\om R_-$ and $-\om R_+$, defined 
in Eqs. (9) and (10) in \cite{josh1}. With these identifications, the 
relation $A=D$ (Eq. (\ref{ADeq})) coincides essentially with Eq. (18) of 
\cite{josh1}. The relations (\ref{BtoCrel}) and (\ref{CtoBrel}) were not 
discussed in \cite{josh1}, but one can verify them, for example, for the 
resolvents corresponding to the kink case $\sigx = m\,{\rm tanh}\, mx$ 
(Eq. (29) in \cite{josh1}), for which $$C=-{\om\over 2\sqrt{m^2-\om^2}}\,,\quad
B=\left[\left({m\, {\rm sech}\, mx\over \om}\right)^2 - 1\right]C\,.$$

\newpage
\subsection{Bilinear Fermion Condensates and Vanishing of the Spatial 
Fermion Current}

Following basic principles of quantum field theory, we may write the most 
generic flavor-singlet bilinear fermion condensate in our static background
as
\beqra\label{condensate}
&&\langle \bar\psi_{a\alpha}(t,x)\,\Gamma_{\alpha\beta}
\,\psi_{a\beta}(t,x)\rangle_{{\rm reg}}
=N\int {d\om\over 2\pi }\,\rmtr\left[\Gamma 
\langle x| {-i\over \om\gam^0 + i\gam^1\pax 
-\left(\si + i\pi\gam_5\right)} | x \rangle_{{\rm reg}}\right]
\nonumber\\{}\nonumber\\ 
&&= N\int {d\om\over 2\pi }\,
\rmtr\left\{\Gamma\left[ 
\left(\begin{array}{cc} A(x) & 
B(x) \\{}&{}\\ C(x) & 
D(x)\end{array}\right)\,-\, \left(\begin{array}{cc} A & 
B \\{}&{}\\ C & 
D\end{array}\right)_{_{VAC}}\right]\right\}\,,
\eeqra
where we have used (\ref{diagonal}).
Here $a=1,\cdots ,N$ is a flavor index, and the trace is taken over 
Dirac indices $\alpha, \beta$. As usual, we regularized this condensate by 
subtracting from it a short distance divergent piece embodied here by the
diagonal resolvent
\beq\label{vac}
\langle x\,|-iD^{-1} | x\,\rangle_{_{VAC}} = 
\left(\begin{array}{cc} A & 
B \\{}&{}\\ C & 
D\end{array}\right)_{_{VAC}} = 
{1\over 2 \sqrt{m^2-\omega^2}}
\left(\begin{array}{cc} i m{\rm cos}\theta & \omega +m{\rm sin}\theta 
\\{}&{}\\ -\omega+m{\rm sin}\theta &  i m{\rm 
cos}\theta\end{array}\right)
\eeq
of the Dirac operator in a vacuum configuration $\si_{_{VAC}} = 
m {\rm cos}\theta$ and $\pi_{_{VAC}} = m{\rm sin}\theta$.

In our convention for $\gam$ matrices (\ref{majorana}) we have 
\beq\label{diagresolvent}
\left(\begin{array}{cc} A(x) & 
B(x) \\{}&{}\\ C(x) & 
D(x)\end{array}\right)\, = {A(x)+D(x)\over 2}{\bf 1} + 
{A(x)-D(x)\over 2i}\gam^1 + i{B(x)-C(x)\over 2}\gam^0 + 
{B(x)+C(x)\over 2}\gam_5\,.
\eeq

An important condensate is the expectation value of the fermion current 
$\langle j^\mu (x)\rangle $. In particular, consider its spatial component.
In our static background $(\sigx ,\pix)$, it 
must, of course, vanish identically
\beq\label{j1vanish}
\langle j^1 (x)\rangle =0\,.
\eeq
Thus, substituting $\Gamma = \gam^1$ in (\ref{condensate}) and using 
(\ref{diagresolvent}) we find 
\beq\label{fourierspat}\langle j^1 (x)\rangle = 
iN\int {d\om\over 2\pi }\,\left[A(x)-D(x)\right]\,.
\eeq
But we have already proved that $A(x)=D(x)$ in {\em any} 
static background $(\sigx ,\pix)$ (Eq.(\ref{ADeq})). Thus, each
frequency component of $\langle j^1 \rangle$ vanishes separately, and 
(\ref{j1vanish}) holds identically. 
It is remarkable that the generalized supersymmetry of the Dirac operator
guarantees the consistency of any static $(\sigx ,\pix)$ 
background.

We discussed $<j^1 (x)> =0$ in \cite{fz}. However, that 
analysis was incomplete as it considered only the asymptotic behavior of 
$<j^1 (x)>$, which misled us to draw an overrestrictive necessary 
consistency condition on the background.

Expressions for other bilinear condensates may be derived in a similar 
manner to the derivation of $<j^1 (x)> $ (here we write the unsubtracted 
quantities). Thus, substituting $\Gamma = \gam^0$ in (\ref{condensate}) and 
using (\ref{diagresolvent}), (\ref{ADeq}) and (\ref{BCfinal}), 
we find that the fermion density is 
\beq\label{fourierdensity}
\langle j^0 (x)\rangle = 
iN\int {d\om\over 2\pi }\,\left[B(x)-C(x)\right] = 
iN\int {d\om\over 2\pi }\,{2\om B(x) -i\pax D(x)\over  \om + \pix}\,.
\eeq
Similarly, the scalar and pseudoscalar condensates are 
\beq\label{fourierscalar}
\langle \bar\psi (x)\psi (x)\rangle = 
N\int {d\om\over 2\pi }\,\left[A(x)+D(x)\right] = 
2N\int {d\om\over 2\pi }\, D(x)\quad\quad \,,
\eeq
and 
\beq\label{fourierpseudoscalar}
\langle \bar\psi (x)\gam^5\psi (x)\rangle = 
N\int {d\om\over 2\pi }\,\left[B(x)+C(x)\right] =
N\int {d\om\over 2\pi }\,{2\pix B(x) +i\pax D(x)\over  \om + \pix}\,.
\eeq

\newpage
\section{The Saddle point Equations and Reflectionless Backgrounds}
\setcounter{equation}{0}
For static backgrounds $(\sigx,\pix)$ we have the (divergent) formal relation
\beqast
\langle x,t |{1\over i\notpa -\left(\si + i\pi\gam_5\right)} | x,t \rangle 
&=& \int{d\om\over 2\pi} \langle x| {1\over \om\gam^0 + i\gam^1\pax 
-\left(\si + i\pi\gam_5\right)} | x \rangle \nonumber\\{}\nonumber\\
&=&i\int{d\om\over 2\pi}\,
\left(\begin{array}{cc} A(x) & B(x) \\{}&{}\\ C(x) & 
D(x)\end{array}\right)
\eeqast
(see Eq. (\ref{condensate})). 
Therefore, using (\ref{majorana}) and (\ref{ADeq}), the bare saddle point 
equations (\ref{saddle}) for static bags are
\beqra
{\del S_{\em eff}\over \del \si\left(x,t\right)}_{|_{\rm static}}  &=&
-{\si\left(x\right)\over g^2} -\,2N~\int{d\om\over 2\pi}\,A(x)= 0
\nonumber\\{}\nonumber\\
{\del S_{\em eff}\over \del \pi\left(x,t\right)}_{|_{\rm static}}  &=&
-{\pi\left(x\right)\over g^2} -\,iN~\int{d\om\over 2\pi}\,\left[B(x)+C(x)\right]= 0\,,
\label{staticsaddle}
\eeqra
where $(\sigx,\pix)$ are subjected to the asymptotic boundary 
conditions (\ref{boundaryconditions}). As was already mentioned following 
(\ref{boundaryconditions}), we further assume that $\sigx$ and $\pix$ tend 
to their asymptotic boundary values on the vacuum manifold at an 
exponential rate which is determined, essentially, by the mass gap $m$ of 
the model, as typical of solitonic configurations.

The $\om$-integrals in (\ref{staticsaddle}) are divergent. 
For bounded bag profiles which satisfy the boundary conditions 
(\ref{boundaryconditions}), the diagonal resolvent 
(\ref{diagonal}) tends, for large $\om$, to that of the vacuum background
(\ref{vac}). Thus, we note from (\ref{vac}), that while 
each of the integrals $\int d\om \,B(x)$ and $\int d\om\, C(x)$ diverges 
linearly with the ultraviolet cutoff, their sum diverges only logarithmically,
as does $\int d\om \,A(x)$. The saddle point equations for vacuum 
condensates, i.e., the gap equations 
\beqra
&&-{\si_{_{VAC}}\over g^2} -\,2N~\int{d\om\over 2\pi}\,A_{
_{VAC}}= 0
\nonumber\\{}\nonumber\\
&&-{\pi_{_{VAC}}\over g^2} -\,iN~\int{d\om\over 2\pi}\,[B_{_{VAC}}+ 
C_{_{VAC}}]= 0\,,
\label{gapsipi}
\eeqra
exhibit the same logarithmic divergence, of course. 
Thus, we can take care of the UV divergence in (\ref{staticsaddle}) by 
subtracting from these equations the corresponding gap equations.

We now concentrate on the subtracted saddle point equation for $\sigx$ 
\beq\label{subtractedsigma}
{\sigx - \si_{_{VAC}} \over 2Ng^2}  = -\,\int_{\cac}
{d\om\over 2\pi}\,\left[A(x) - A_{_{VAC}}\right]\,.
\eeq
The integration contour ${\cal C}$ in (\ref{subtractedsigma}) is 
commonly\footnote{See e.g., Section 4 of \cite{dhn1} for a detailed 
discussion. If the bound state at energy $\om_1$ traps $n_f$ 
fermions, then routing the contour ${\cal C}$ as indicated in Fig. (1) means 
that we are discussing the sector of the model with fermion number $n_f$.}
taken as indicated in Fig.(1), which shows qualitatively the spectrum of the 
Dirac equation in a bag background. 
\vspace{24pt}
\par
\hspace{0.5in} \epsfbox{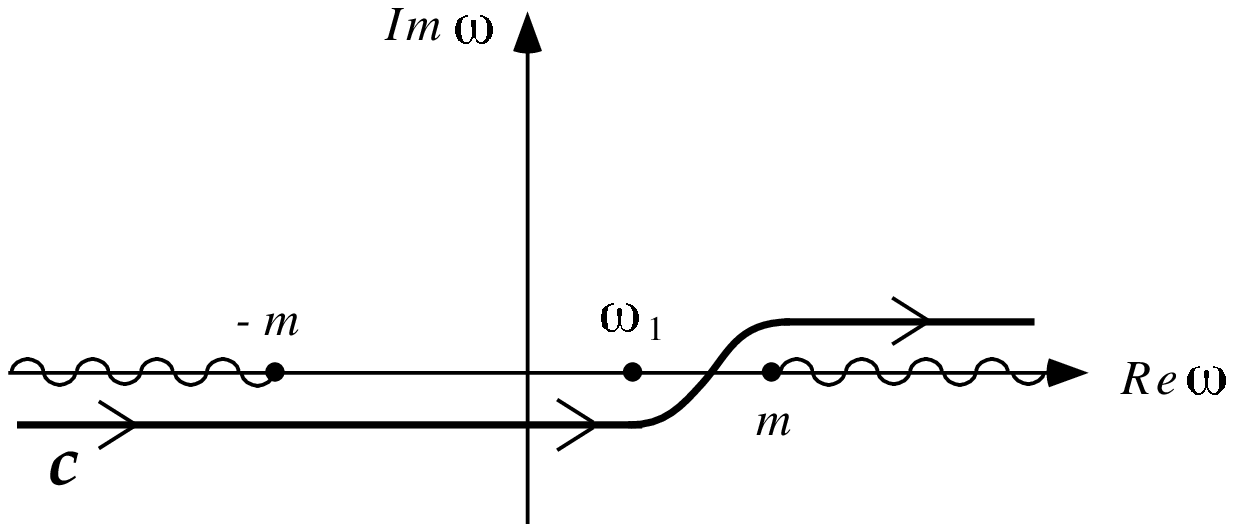}
\par
\baselineskip 12pt
{\footnotesize {\bf Fig.~1:} The contour ${\cal C}$ in the complex $\om$
plane in Eq. (\ref{subtractedsigma}). The continuum states appear as the 
two cuts along the real axis with branch points at $\pm m$ (wiggly lines), 
and the bound state is the pole at $\om_1$.}
\vspace{24pt}
\par
\baselineskip 20pt   
Besides the continuum states in that 
spectrum (the two cuts corresponding to the Fermi sea of negative energy 
states $\om\leq -m$, and scattering states with $\om\geq m$), there are 
bound states within the gap 
$-m\leq\om\leq m$, which trap fermions 
into the bag. One such bound state is indicated in Fig. (1) as the pole 
at $\om=\om_1$. The detailed calculation to determine bound state energies 
like $\om_1$ is discussed in \cite{shei, fz} (after establishing the 
reflectionless property of the background). We stress that 
$\si_{_{VAC}}$ in (\ref{subtractedsigma}) can 
be the $\si$ component of {\em any} point on the vacuum manifold 
$\si^2 +\pi^2 = m^2$ (and similarly for $\pi_{_{VAC}}$, which appears in the 
subtracted equation for $\pix$).

The contour integral in (\ref{subtractedsigma}) is most conveniently 
calculated by deforming the contour ${\cal C}$ into the contour ${\cal C}'$ 
shown in Fig.(2). (This is allowed, since the subtracted integral in 
(\ref{subtractedsigma}) converges.) The ``hairpin" wing of ${\cal C}'$ picks 
up the contribution of the filled Fermi sea, and the little circle 
around the simple pole at $\om=\om_1$ is the contribution of fermions 
populating the bound state of the ``bag".
\par
\vspace{24pt}
\hspace{0.5in} \epsfbox{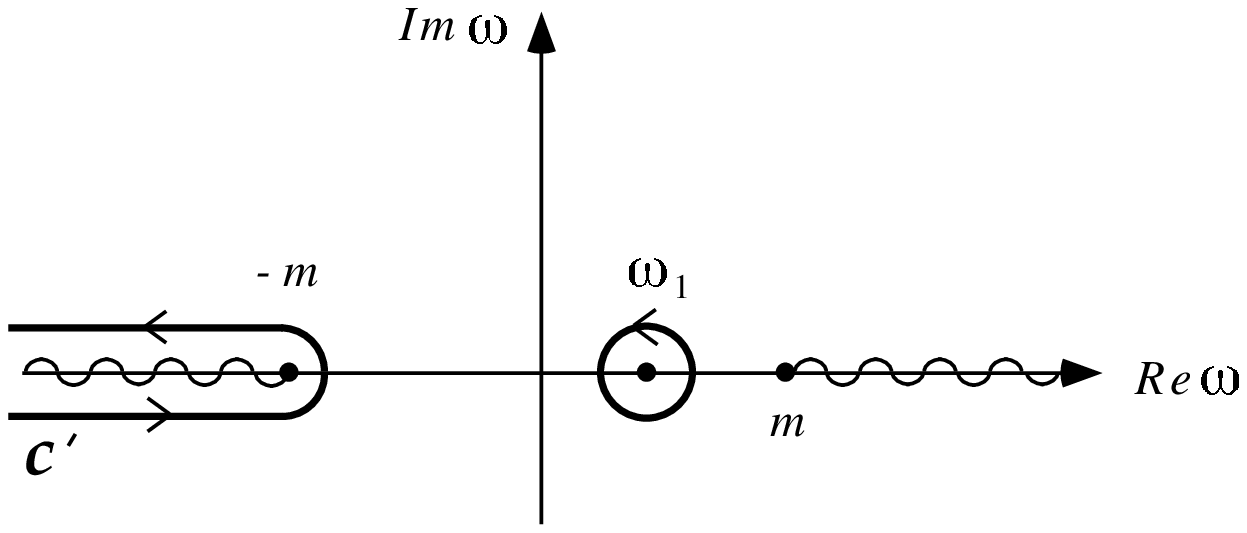}
\par
\baselineskip 12pt
{\footnotesize {\bf Fig.~2:} The deformed integration contour ${\cal C}'$
used in Eq. (\ref{subtractedsigma}).}
\vspace{24pt}
\par
\baselineskip 20pt
Let us study the spatial asymptotic behavior of (\ref{subtractedsigma}). 
From (\ref{condensateasp}) we have 
\beqra\label{Aas}
A(x) - A_{_{VAC}}(-\infty)&&\asymptoticm {ik - \si (-\infty)\over 2k} 
\, r_2(k) e^{-2ikx}
\nonumber\\{}\nonumber\\
A(x) - A_{_{VAC}}(+\infty)&&\asymptoticp {ik + \si (+\infty)\over -2k}
\, r_1(k) e^{2ikx}\,,
\eeqra
where $\si (\pm\infty)$ are the appropriate vacuum boundary values of $\si$ 
from  (\ref{boundaryconditions}), and $r_1(k), r_2(k)$ are the 
reflection coefficients defined in (\ref{otherside}).  

Thus, for example, studying (\ref{subtractedsigma}) as $x\rightarrow\infty$, 
we see from (\ref{Aas}) that 
\beq\label{sigas}
{\sigx - \si (+\infty) \over 2Ng^2} \asymptoticp  
\int_{\rm Fermi~sea}{d\om\over 2\pi}\,{ik + \si (+\infty)\over 2k}
\, r_1(k) e^{2ikx} +\cO (e^{-{\rm const.}\,mx})\,.
\eeq
The first term in (\ref{sigas}) is the contribution coming from the Fermi 
sea (i.e., the ``hairpin" wing of ${\cal C}'$). The second, 
exponentially small term on the right hand side of (\ref{sigas}) comes from 
the bound state pole (i.e., it is proportional to the bound state 
wave function squared). Due to the asymptotic boundary conditions on $\sigx$,
the left hand side of (\ref{sigas}) is also exponentially small as 
$x\rightarrow\infty$. Thus, the first term on the right hand side of
(\ref{sigas}) must have an exponentially small bound as $x\rightarrow\infty$.

We now change the variable to $k=\sqrt{\om^2-m^2}$. When mapping into the 
$k$-plane, the lower wing
of the cut in Fig.(2) is transformed into $k=|k|e^{i\delta}$, and the upper 
wing is transformed into 
$-|k|e^{-i\delta}$, with $\delta\rightarrow 0+$\footnote{For 
example, just above the cut $\om-m=|\om-m|e^{i(\pi-\delta)}$ and 
$\om+m=|\om+m|e^{i(\pi-\delta)}$, with $\delta\rightarrow 0+$. Thus,
just above the cut, $k=\sqrt{\om^2-m^2}=\sqrt{|\om^2-m^2|}e^{i(\pi-\delta)}
=-|k|e^{-i\delta}$.}. 
Thus, we may write the dispersion integral on the right hand side of 
(\ref{sigas}) coming from states in the Fermi sea as $-\Delta (x)$, where
\beqra\label{dispersion}
\Delta (x) &=&
\int\limits_0^\infty {dk\over 4\pi}\,
{(ik+\si(\infty))\, r_1(k) e^{2ikx}
+(k\rightarrow -k)\over \sqrt{k^2+m^2}}\nonumber\\{}\nonumber\\
&=& {\rm Re}\left[\int\limits_0^\infty {dk\over 2\pi}\,
{(ik+\si(\infty))\, r_1(k) e^{2ikx}\over \sqrt{k^2+m^2}}\right]\,,
\eeqra
where in the last equality we used the reflection property 
$r_1(-k) = r_1^*(k)$ (\ref{reflectionproperty}) of $r_1(k)$.

The function $\Delta (x)$ has to die off at least at an exponential rate 
as $x\rightarrow\infty$. Thus, we are to study the asymptotic 
behavior of 
\beq\label{gx}
G (x) = \int\limits_0^\infty {dk\over 2\pi}\,
{(ik+\si(\infty))\, r_1(k) e^{2ikx}\over \sqrt{k^2+m^2}}
\eeq
at large $x$. To this end we have to invoke some of the 
general properties of the reflection coefficient $r_1(k)$ of the operator 
$L_b$ in (\ref{bcops}).

Due to the boundary conditions (\ref{boundaryconditions}) on 
the background fields $\sigx$ and $\pix$, the operator $L_b$ tends 
exponentially fast (in $x$) to its asymptotic free particle form. 
Thus, its ``scattering potential'' is localized in a finite region in space.

From the literature on scattering theory (in one space dimension)
\cite{faddeev} we know that the reflection coefficient $r(k)$ of 
Schr\"odinger operators with short range potential wells\footnote{We tacitly 
assume that the potential wells in question tend to the same asymptotic 
value at $x=\pm\infty$ (as $L_b$ does with $\sigx, \pix$ satisfying 
(\ref{boundaryconditions})), and that they do not have any barriers above 
these asymptotic values.} is analytic on the real $k$ axis (and generally 
follows the threshold behavior $r(k)= -1 + ak +\cdots$) and dies off 
like $1/k$ as $k\rightarrow\infty$, i.e., at large kinetic energy.
Strictly speaking, the discussion of these issues in the various 
references in \cite{faddeev} concentrates mostly on Schr\"odinger operators of 
the standard form $-\pa_x^2 + V(x)$, but the arguments leading to the 
conclusions about the behavior of $r(k)$ may be easily generalized to the 
scattering theory of the Sturm-Liouville operators $L_b$ and $L_c$ in 
(\ref{bcops}) (with $\sigx$ and $\pix$ relaxing fast to 
(\ref{boundaryconditions})).

Therefore, in deriving the asymptotic behavior of $G(x)$ and 
$\Delta (x)$ we may use as 
an input that $r_1(k)$ is analytic on the real $k$ axis and that it decays 
at least as fast as $1/k$ as $k\rightarrow\infty$. Given these properties 
of $r_1(k)$, we are allowed to expand $G (x)$ in powers of $1/x$ in the 
most natural way, namely, by repeatedly integrating 
by parts over $k$ in (\ref{gx}).

Thus, for example, after three integrations we find  
\beqra\label{intbyparts}
2\pi G (x) &=&
-{\si (\infty)r_1(0)\over 2imx} 
+ {imr_1(0) +m\si (\infty)r_1'(0)\over (2imx)^2}\nonumber\\{}\nonumber\\
&-&{-\si (\infty)r_1(0) +2im^2 r_1'(0) + m^2\si (\infty) r_1''(0)
\over (2imx)^3}\nonumber\\{}\nonumber\\
&-&\int\limits_0^\infty dk {e^{2ikx}\over (2ix)^3}{\pa^3\over\pa k^3}
\left({ik+\si (\infty)\over\sqrt{k^2 + m^2}}\,r_1(k)\right)\,,
\eeqra
and so on and so forth. Clearly, the remaining integral in each step is 
subdominant by a power of $1/x$ relative to its predecessor, and thus, the 
expansion of $G (x)$ generated in this way is an asymptotic expansion.

From (\ref{intbyparts}) (or by working out a few more terms in the 
asymptotic expansion if necessary) the following pattern emerges: 
the coefficient of $(1/2ix)^{2n+1}$ is a linear combination of the form
$$\sum_{j=0}^{2n} c_j i^j r_1^{(j)}(0)$$ with {\em real} coefficients $c_j$, 
and the coefficient of $(1/2ix)^{2n}$ is a linear combination of the form
$$i\sum_{j=0}^{2n-1} c_j i^j r_1^{(j)}(0)$$ with some other real 
coefficients $c_j$.

From the reflection property (\ref{reflectionproperty}) $r_1(-k) = r_1^*(k)$
we immediately conclude that $r_1^{(n)} (0) = (-1)^n r_1^{(n)}(0)^*$: 
the even derivatives $r_1^{(2j)}(0)$ are real, and the odd derivatives 
$r_1^{(2j+1)}(0)$ are imaginary. Thus, the linear combinations 
$\sum c_j i^j r_1^{(j)}(0)$ are real, which makes
each term on the right hand side of the asymptotic expansion 
(\ref{intbyparts}) of $G(x)$ {\em pure imaginary}.

Using this result in (\ref{dispersion}) we conclude that all terms in 
the asymptotic expansion of $\Delta (x)$ in powers of $1/x$ vanish. 
Thus, $\Delta (x)$ vanishes faster than any power of $1/x$ as
$x\rightarrow\infty$. This is consistent with our expectation that 
$\Delta (x)$ vanishes at least at an exponential rate when 
$x\rightarrow\infty$.

This concludes our discussion of the subtracted 
saddle point equation (\ref{subtractedsigma}) for $\sigx$ and its 
asymptotic behavior. 

We can repeat the same story 
for the subtracted saddle point equation for $\pix$ 
\beq\label{subtractedpi}
{\pi\left(x\right)-\pi_{_{VAC}}\over iNg^2} = -
\int{d\om\over 2\pi}\,\left[\left(B(x)-B_{_{VAC}}\right)+
\left(C(x)-C_{_{VAC}}\right)\right]\,.
\eeq
In a similar manner to our derivation of (\ref{sigas}) and 
(\ref{dispersion}), we can show that 
\beqra\label{pias}
{\pix - \pi (\infty) \over Ng^2}\asymptoticp {\rm exponentially~small~
contribution~of~bound~states} \nonumber\\{}\nonumber\\-{\rm Re}\left[
\int\limits_0^\infty {dk\over 2\pi}\,
{(\sqrt{k^2+m^2}-\pi(\infty)) -(\sqrt{k^2+m^2}+\pi(\infty)) {
\si(\infty)+ik\over \si(\infty)-ik}\over \sqrt{k^2+m^2}}\, 
r_1(k) e^{2ikx}
\right]\,.
\eeqra
Due to the asymptotic boundary conditions on $\pix$,
the left hand side of (\ref{pias}) is exponentially small as 
$x\rightarrow\infty$, which bounds the dispersion integral on the right 
hand side of (\ref{pias}). As in our analysis of (\ref{gx}), we expand 
the integral in the square brackets in (\ref{pias}) in powers of $1/x$, and 
similarly to (\ref{intbyparts}), we can show that all terms in that 
asymptotic series are pure imaginary. Thus, the right hand side of 
(\ref{pias}) vanishes faster than any power of $1/x$, consistent with the
boundary conditions on $\pix$.

We conclude that the asymptotic behavior of the static saddle point equations 
(\ref{subtractedsigma}) and (\ref{subtractedpi}) is consistent with
the asymptotic boundary conditions (\ref{boundaryconditions}) on $\sigx$ and 
$\pix$ for {\em any} reflection amplitude $r_1 (k)$: all terms in the 
asymptotic expansions of the dispersion integral
in the square brackets in (\ref{pias}) and also of $G(x)$ in (\ref{gx}) 
in powers of $1/x$ are imaginary.

Contrary to the argument we made in \cite{fz}, the reflectionless property of 
the solutions $\sigx$ and $\pix$ of (\ref{staticsaddle}) does not emerge 
as a necessary condition from consistency of the asymptotic behavior of 
(\ref{sigas}) and (\ref{pias}) and the boundary conditions on the 
background fileds.


\newpage
\setcounter{equation}{0}
\setcounter{section}{0}
\renewcommand{\theequation}{A.\arabic{equation}}
\renewcommand{\thesection}{Appendix:}
\section{Asymptotics of the Dirac Resolvent}
\vskip 5mm
\setcounter{section}{0}
\renewcommand{\thesection}{A}

In this Appendix we discuss the spatial asymptotic behavior of the 
diagonal resolvent of the Dirac operator (\ref{diagonal}). 

According to our discussion in Section 2.2, given, for example, $B(x)$, 
we may determine $D(x)$, $A(x)$ and $C(x)$ using (\ref{abcd}) first, then 
(\ref{ADeq}) and finally, (\ref{BtoCrel}). 

Thus, it is enough to determine the asymptotic behavior of $B(x)$. According 
to (\ref{abcd}), $B(x)=b_1(x)b_2(x)/W_b$, so we need to determine the 
asymptotic behavior of $b_1(x), b_2(x)$, the fundamental solutions of 
$L_bb(x)=0$.

By definition, according to (\ref{planewaves}), and with the standard choice
(\ref{standardchoice}) of the coefficients 
$A_{b}^{(1)} = A_{b}^{(2)} =1$, these functions satisfy
$$b_1(x) \asymptoticm e^{-ikx} \quad\quad ,\quad\quad 
b_2(x)\asymptoticp e^{ikx}$$
at the two opposite sides of the one dimensional world. Thus, what remains is 
to determine the asymptotic behavior of each of these functions on the 
other side of the world. Since $\si$ and $\pi$ relax to the vacuum manifold
as $x\rightarrow\pm\infty$ (Eq. (\ref{boundaryconditions})), the operators 
$L_b$ and $L_c$ in (\ref{bcops}) degenerate into free particle operators, 
and thus we must have 
\beqra\label{otherside}
b_1(x) &&\asymptoticp {1\over t_1(k)}e^{-ikx} + {r_1(k)
\over t_1(k)}e^{ikx}\nonumber\\{}\nonumber\\
b_2(x) &&\asymptoticm {1\over t_2(k)}e^{ikx} + {r_2(k)
\over t_2(k)}e^{-ikx}\,,
\eeqra
where $t_{1,2}(k), r_{1,2}(k)$ are the appropriate transmission and reflection
amplitudes, respectively. 

It is enough to consider $k\geq 0$, since the operators $L_b$ and $L_c$ in 
(\ref{bcops}) are real and thus $b_{1,2} (x, -k) = b_{1,2}^*(x,k)$, leading 
to 
\beq\label{reflectionproperty}
t_{1,2}(-k) = t_{1,2}^* (k)\quad {\rm and}\quad r_{1,2}(-k) = r_{1,2}^* (k)\,.
\eeq
Thus, $b_1(x)$ corresponds to a setting with 
a source at $x=+\infty$ which emits to the left, and $b_2(x)$ describes a 
source at $x=-\infty$ which emits to the right.

The wronskian of $b_1(x)$ and $b_2(x)$ (Eq. (\ref{wronskian}))
$$W_b(\om) ={b_2(x)b_1^{'}(x)-b_1(x)b_2^{'}(x)\over \omega+\pix}$$
is independent of $x$. Thus, evaluating it at $x\rightarrow\pm\infty$ we find
\beq\label{wb}
W_b(\om) = {-2ik\over t_2(k) (\om + \pi(-\infty))} = 
{-2ik\over t_1(k) (\om + \pi(+\infty))}\,,
\eeq
and thus, 
\beq\label{t1t2}
t_1(k)(\om + \pi(+\infty)) =  t_2(k) (\om + \pi(-\infty))\,.
\eeq

Like the wronskian of $\{b_1(x), b_2(x)\}$, the wronkians of the  
pairs of independent solutions $\{b_1(x), b_1^*(x)\}$ and 
$\{b_2(x), b_2^*(x)\}$ (here we assume that $k$ is real)
are also independent of $x$. In fact, these wronskians are proportional to 
the Schr\"odinger probability currents carried by $b_1(x)$ and $b_2(x)$, 
respectively. 
Thus, using (\ref{planewaves}) and (\ref{otherside}) to evaluate each of these 
wronskians at $x=\pm\infty$ and then equating the results, we obtain
\beqra\label{w1w2}
W_b [b_1, b_1^*] &=& {b_1^*(x)b_1^{'}(x)-b_1(x)b_1^{*'}(x)\over \omega+\pix}
\nonumber\\{}\nonumber\\
&=& {-2ik\over \om + \pi(-\infty)} = 
{-2ik\over \om + \pi(+\infty)} {1-|r_1|^2\over |t_1|^2}\,;
\nonumber\\{}\nonumber\\
W_b [b_2, b_2^*] &=& {b_2^*(x)b_2^{'}(x)-b_2(x)b_2^{*'}(x)\over \omega+\pix}
\nonumber\\{}\nonumber\\
&=& {2ik\over \om + \pi(+\infty)} = 
{2ik\over \om + \pi(-\infty)} {1-|r_2|^2\over |t_2|^2}\,.
\eeqra
Thus, we deduce the generalized unitarity conditions
\beq\label{tratio}
{1-|r_1|^2\over |t_1|^2} = {|t_2|^2 \over 1-|r_2|^2} = 
{\om + \pi(+\infty)\over \om + \pi(-\infty)}\,.
\eeq
The usual unitarity condition $|t|^2 + |r|^2 =1$ holds for backgrounds in
which $\pi(+\infty) = \pi(-\infty)$.

Putting every thing together, we finally learn that 
\beqra\label{Bas1}
B(x)&&\asymptoticm {1+r_2(k) e^{-2ikx}\over -2ik} (\om + \pi(-\infty))
\nonumber\\{}\nonumber\\
B(x)&&\asymptoticp {1+r_1(k) e^{2ikx}\over -2ik} (\om + \pi(+\infty))\,.
\eeqra

Recall from (\ref{vac}) that 
\beq\label{Bvacas}
B_{_{VAC}} (\pm\infty) = {\om + \pi(\pm\infty)\over 2\sqrt{m^2-\om^2}}
= {\om + \pi(\pm\infty)\over -2ik}
\eeq
corresponds to the appropriate vacuum configurations $(\si (\pm\infty), 
\pi (\pm\infty))$. Thus, we can rewrite (\ref{Bas1}) as 
\beqra\label{Bas}
B(x)&&\asymptoticm B_{_{VAC}} (-\infty) \left(1+r_2(k) e^{-2ikx}\right) 
\nonumber\\{}\nonumber\\
B(x)&&\asymptoticp B_{_{VAC}} (+\infty)\left( 1+r_1(k) e^{2ikx}\right) \,.
\eeqra

Using (\ref{ADeq}), (\ref{cfromb}) and (\ref{BtoCrel}) we then find the 
asymptotic behaviors of the remaining entries of the diagonal resolvent 
(\ref{diagonal}): 
\beqra\label{ADas}
A(x) = D(x)&&\asymptoticm D_{_{VAC}} (-\infty) + 
{ik-\si (-\infty)\over 2k}\, r_2(k) e^{-2ikx} 
\nonumber\\{}\nonumber\\
A(x) = D(x)&&\asymptoticp D_{_{VAC}} (+\infty) - 
{ik + \si (+\infty)\over 2k}\, r_1(k) e^{2ikx} \,,
\eeqra
and 
\beqra\label{Cas}
C(x)&&\asymptoticm C_{_{VAC}} (-\infty)\left[ 1+ {\si(-\infty) -ik\over
\si (-\infty) +ik }\, r_2(k) e^{-2ikx}\right] 
\nonumber\\{}\nonumber\\
C(x)&&\asymptoticp C_{_{VAC}} (+\infty)\left[ 1+ {\si(+\infty) +ik\over
\si (+\infty) -ik }\, r_1(k) e^{2ikx}\right] \,,
\eeqra
with \beq\label{Cvacas}
C_{_{VAC}} (\pm\infty) = {-\om + \pi(\pm\infty)\over 2\sqrt{m^2-\om^2}}
= {\om - \pi(\pm\infty)\over 2ik}
\eeq
from (\ref{vac}).

Finally, we can write these results more compactly as 
\beqra\label{condensateasp}
&&\left(\begin{array}{cc} A(x) & 
B(x) \\{}&{}\\ C(x) & 
D(x)\end{array}\right)\,\asymptoticp\, \left(\begin{array}{cc} A(+\infty) & 
B(+\infty) \\{}&{}\\ C(+\infty) & 
D(+\infty)\end{array}\right)_{_{VAC}}\,+\nonumber\\{}\nonumber\\{}
\nonumber\\
&&\left(\begin{array}{cc} {ik + \si (+\infty)\over -2k} & 
B_{_{VAC}}(+\infty) \\{}&{}\\ C_{_{VAC}}(+\infty) 
{\si(+\infty) +ik\over\si (+\infty) -ik }& 
{ik + \si (+\infty)\over -2k}\end{array}\right)\, r_1(k) e^{2ikx}\,,
\nonumber\\{}\nonumber\\{}
\nonumber\\
&&\left(\begin{array}{cc} A(x) & 
B(x) \\{}&{}\\ C(x) & 
D(x)\end{array}\right)\,\asymptoticm\, \left(\begin{array}{cc} A(-\infty) & 
B(-\infty) \\{}&{}\\ C(-\infty) & 
D(-\infty)\end{array}\right)_{_{VAC}}\,+\nonumber\\{}\nonumber\\{}
\nonumber\\
&&
\left(\begin{array}{cc} {ik - \si (-\infty)\over 2k}& 
B_{_{VAC}}(-\infty)\\{}&{}\\ C_{_{VAC}}(-\infty) 
{\si(-\infty) +ik\over\si (-\infty) -ik }& 
{ik - \si (-\infty)\over 2k}\end{array}\right)\, r_2(k) e^{-2ikx}\,.
\eeqra

\vspace{2cm}
{\bf Acknowledgements}~~~ We would like to thank R. Jaffe and N. Graham 
for useful correspondence and for pointing out a problem 
in the ``proof'' of reflectionless in \cite{fz}. J.F. thanks M. Moshe 
and J. Avron for useful discussions. The work of A.Z. was supported in 
part by the National Science Foundation under Grant No. PHY 94-07194. 
J.F.'s research 
has been supported in part by the Israeli Science Foundation grant number 
307/98 (090-903).


\newpage
\begin{thebibliography}{99}
\bibitem{dhn}  R.F. Dashen, B. Hasslacher and A. Neveu,  Phys. Rev. D 
{\bf 12}, 2443 (1975).

\bibitem{shei} S. Shei,  Phys. Rev. D {\bf 14}, 535 (1976).

\bibitem{faddeev}  L.D. Faddeev and L.A. Takhtajan, {\sl Hamiltonian 
Methods
in the Theory of Solitons\/} (Springer Verlag, Berlin, 1987).\\
L.D. Faddeev, J. Sov. Math. {\bf 5} (1976) 334. This paper is reprinted 
in\newline
L.D. Faddeev, {\sl 40 Years in Mathematical Physics} (World Scientific, 
Singapore, 1995). In section 2.2 of that paper there is a brief discussion
of some aspects of the inverse scattering problem for the Dirac equation 
(\ref{diraceq}). \\   
S. Novikov, S.V. Manakov, L.P. Pitaevsky and V.E. Zakharov, {\sl Theory of 
Solitons - The Inverse Scattering Method\/} 
(Consultants Bureau, New York, 1984)\newline
( Contemporary Soviet Mathematics). 

\bibitem{jg} J. Goldstone and R. Jackiw, Phys. Rev. D  {\bf 11}, 1486 
(1975).

\bibitem{sphericalbag} T. D. Lee and G. Wick, Phys. Rev. D {\bf 9}, 2291 
(1974);\\
R. Friedberg, T.D. Lee and R. Sirlin, Phys. Rev. D {\bf 13}, 2739 
(1976);\\
R. Friedberg and T.D. Lee, Phys. Rev. D {\bf 15}, 1694 (1976), {\em 
ibid.} 
{\bf 16}, 1096 (1977);\\
A. Chodos, R. Jaffe, K. Johnson, C. Thorn, and V. Weisskopf,  Phys. Rev. 
D {\bf 9}, 3471 (1974).

\bibitem{shellbag} W. A. Bardeen, M. S. Chanowitz, S. D. Drell, M. 
Weinstein 
and T. M. Yan, Phys. Rev. D {\bf 11}, 1094 (1974);\\
M. Creutz, Phys. Rev. D {\bf 10}, 1749 (1974).


\bibitem{gn}  D.J. Gross and A. Neveu,  Phys. Rev. D  {\bf 10},   3235
(1974).

\bibitem{njl} Y. Nambu and G. Jona-Lasinio, Phys. Rev. {\bf 122}, 345 
(1961), {\it ibid} {\bf 124}, 246 (1961).

\bibitem{ccgz} C.G. Callan, S. Coleman, D.J. Gross and A. Zee,
unpublished; This work is described by 
D.J. Gross in {\sl Methods in Field Theory\/}, R. Balian and J. 
Zinn-Justin (Eds.), Les-Houches session  XXVIII 1975 (North Holland, 
Amsterdam, 1976).

\bibitem{rosner} 
H. B. Thacker, C. Quigg and J. L. Rosner, Phys. Rev. D {\bf 18}, 274 (1978);\\
J. F. Schonfeld, W. Kwong, J. L. Rosner, C. Quigg and H. B. Thacker, 
Ann. Phys. {\bf 128}, 1 (1980);\\  
W. Kwong and J. L. Rosner, Prog. Theor. Phys. (Suppl.) {\bf 86}, 366 (1986);\\ 
J. L. Rosner, Ann. Phys. {\bf 200}, 101 (1990);\\
A. K. Grant and J. L. Rosner, Jour. Math. Phys. {\bf 35}, 2142 (1994), 
and references therein.
 
\bibitem{blackhole} S. Chandrasekhar, {\sl The Mathematical Theory of 
Black Holes\/}, chapter 4 (Oxford University Press, New York, 1992).  

\bibitem{zwiebach} B. Zwiebach, JHEP {\bf 0009}:028,2000 (hep-th/0008227);
J. A. Minahan and B. Zwiebach, JHEP {\bf 0009}:029,2000 (hep-th/0008231), 
JHEP {\bf 0103}:038,2001 (hep-th/0009246), JHEP {\bf 0102}:034,2001 
(hep-th/0011226).

\bibitem{cooper} F. Cooper, A. Khare and U. Sukhatme, Phys. Rep. {\bf 251}, 
267 (1995).

\bibitem{others} See e.g., A. Klein, Phys. Rev. D {\bf 14}, 558 (1976);\\
R. Pausch, M. Thies and V. L. Dolman, Z. Phys. A {\bf 338}, 441 (1991).

\bibitem{thies} V. Schoen and M. Thies, {\sl 2d Model Field Theories at 
Finite Temperature and Density}, Contribution to the Festschrift in honor of 
Boris Ioffe, (M. Shifman Ed.), hep-th/0008175.

\bibitem{jaffe} E. Farhi, N. Graham, R.L. Jaffe and H. Weigel, Nucl. Phys. 
{\bf B585}, 443 (2000);  Phys. Lett. {\bf B475}, 335 (2000).

\bibitem{bashinsky} S. V. Bashinsky, Phys. Rev. D {\bf 61}, 105003 (2000).
 
\bibitem{gd} I.M. Gel'fand  and L.A. Dikii, Russian Math. Surveys~~{\bf
30},~77
{}~(1975).

\bibitem{josh1} J. Feinberg,  Phys. Rev. D {\bf 51}, 4503 (1995).

\bibitem{josh2} J. Feinberg, Nucl. Phys. {\bf B433}, 625 (1995).

\bibitem{fz}  J. Feinberg and A. Zee,  Phys. Rev. D 
{\bf 56},5050 (1997); Int. J. Mod. Phys. {\bf A12}, 1133 (1997). 

\bibitem{massivegn} J. Feinberg and A. Zee,  Phys. Lett. 
{\bf B411}, 134 (1997)

\bibitem{stone} I. Kosztin, S. Kos, M. Stone and A. J. Leggett, 
Phys. Rev. B {\bf 58}, 9365 (1998); S. Kos and M. Stone, Phys. Rev. B 
{\bf 59}, 9545 (1999).

\bibitem{generalizedsusy} J. Feinberg\, ``The Dirac Operator in a Fermion 
Bag Background in $1+1$ Dimensions and Generalized Supersymmetric 
Quantum Mechanics'', hep-th/0109043.


\bibitem{decouple}
See the concluding remarks in \cite{shei} who brings an unpublished 
argument
for decoupling of $\theta$ due to R. Dashen, along the lines of\\
M.B. Halpern, Phys. Rev. D {\bf 12} 1684 (1975).\\ 
See also E. Witten,  Nucl. Phys. {\bf B145}, 110 (1978).

\bibitem{coleman} N. D. Mermin and H. Wagner, Phys. Rev. Lett. 
{\bf 17}, 1133 (1966);\\ 
S. Coleman, Commun. Math. Phys. {\bf 31}, 259 (1973). 

\bibitem{cjt} J.M. Cornwal, R. Jackiw, and E. Tomboulis, Phys. Rev. D  
{\bf 10}, 2428 (1974).

\bibitem{bh} S. Hikami and E. Br\'ezin, J. Phys. A (Math. Gen.) {\bf 12}, 
759 (1979).

\bibitem{dev} H.J. de Vega, Commun. Math. Phys. {\bf 70}, 29 (1979);\\
J. Avan and H.J. de Vega,  Phys. Rev. D  {\bf 29}, 2891 and 1904
(1984)


\bibitem{mackenzie} R. MacKenzie, F. Wilczek and A.Zee,  Phys. Rev. Lett 
{\bf 53}, 2203 (1984).


\bibitem{fd} G. Dunne and J. Feinberg, Phys. Rev. D {\bf 57}, 1271 (1998).

\bibitem{dhn1}  R.F. Dashen, B. Hasslacher and A. Neveu,  Phys. Rev. D 
{\bf 10}, 4130 (1974).

\end{thebibliography}
\end{document}